# High-Pressure Induced Structural Phase Transition in $CaCrO_4$: Evidence from Raman Scattering Studies

Y. W. Long, W. W. Zhang, L. X. Yang, Y. Yong, R. C. Yu, S. Ding, Y. L. Liu, C. Q. Jin[*]

Institute of Physics, Chinese Academy of Sciences, Beijing 100080, P. R. China

## Abstract

Raman spectroscopic studies have been carried out on $CaCrO_4$ under pressure up to 26GPa at ambient temperature. The Raman spectra showed $CaCrO_4$ experienced a continuous structural phase transition started at near 6GPa, and finished at about 10GPa. It is found that the high-pressure phase could be quenched to ambient conditions. Pressure dependence of the Raman peaks suggested there existed four pressure regions related to different structural characters. We discussed these characters and inferred that the nonreversible structural transition in $CaCrO_4$, most likely was from a zircon-type ($I4_1/amd$) ambient phase to a scheelite-type high pressure structure ($I4_1/a$).

$CaCrO_4$ is an important cathode material in thermally activated

electrochemical cells, and it also often be used in many industrial process as an intermediate compound formed in the system of Ca-Cr-O[1-3]. The thermodynamic properties of $CaCrO_4$, such as enthalpy, entropy, Gibbs free energy, heat capacity and so on have been investigated in detail for the wide application of $CaCrO_4$[4-8]. Temperature, chemical potential and pressure are the three fundamental parameters to govern the thermodynamic state of materials. Traditionally temperature and chemical composition are the routine concerns of conventional materials. However with the significant progress of high pressure technique, especially the breakthrough of integrated experimental methods based on diamond anvil cell (DAC), high pressure is gradually becoming a reachable physical dimension in modifying materials state. $CaCrO_4$ was conventionally regarded as a thermally stable material at the temperature even higher than $1000^{o}C$, and its structural properties under high pressure were never reported in the literature as yet to our knowledge. It is known that Raman spectroscopy has been proved to be a convenient means to study structural characters of materials through the solid-state effects on the dynamical properties. Due to the development of DAC methods, high-pressure Raman spectroscopy has been available to investigate materials of pressure-related structural characters[9-10]. We present here for the first time a pressure-dependent Raman scattering study to detect the structural characters of $CaCrO_4$ under high pressure. We found the evidence of pressure-induced structural transition at 6GPa, and analyzed different structural characters in four pressure regions. Furthermore, we concluded the most likely structural transition route was from zircon-type to scheelite-type structure taking account of the available reports on other similar compounds[11-14].

$CaCrO_4$ crystallizes into a zircon-type structure of tetragonal lattice with space group $I4_1/amd$ (Z=4) at ambient conditions. The structure of zircon-type $CaCrO_4$ is built from isolated and slightly distorted $CrO_4$ tetrahedral units, which are connected with $CaO_8$ dodecahedra by edge- and corner-sharing. The detailed structural description was discussed in Ref. 15 and 16. The high pure $CaCrO_4$ polycrystalline powder used in this experiment was synthesized by conventional solid-state reaction method[16]. The powder sample was carefully loaded into a DAC with 500μ m culets. T301 stainless steel with a 250μ m hole and 300μ m thickness was used as gasket. 4:1 methanol-ethanol mixture was used as pressure medium. The pressure was calibrated by the ruby luminescence method[17-18]. The Raman spectra and the ruby luminescence were collected in the back-scattering geometry using a micro-Raman spectrometer (Jobin Yvon T64000) equipped charge-couple device (CCD) detector. A 25x microscope objective lens was used for focusing the laser beam and collection of the scattered light. The source for excitation was Verdi-2 solid-state laser (532 nm), and instrument resolution was 1 $cm^{-1}$.

Based on the point group analysis, we chose the phonons frequency range from 150 to 1000cm$^{-1}$ to detect the internal vibration of $CrO_4$ tetrahedra in $CaCrO_4$[19]. All of the Raman spectra in various pressures up to 26GPa were shown in Figure 1 and 2. Clearly the Raman peaks under different pressure are assigned in two frequency regions: 200-500cm$^{-1}$ and 800-1000cm$^{-1}$, corresponding to O-Cr-O bending modes and Cr-O stretching modes in $CrO_4$ tetrahedra, respectively. At ambient pressure, six unambiguous Raman peaks were observed as reported in Ref. 20. According to the description of Ref. 19, the six Raman peaks were identified as $\nu_1$ ($A_g$), (symmetric-stretching modes), $\nu_3$ ($B_g$ and $E_g$), (antisymmetric-stretching modes), $\nu_2$ ($A_g$ and $B_g$), (symmetric-bending modes), and $\nu_4$ ($B_g$ and $E_g$), (antisymmetric-bending modes). The general rule is that $\nu_1$ and $\nu_3$ modes appear in a higher frequency region than $\nu_2$ and $\nu_4$ modes, and the intensity of the $\nu_1$ ($A_g$) peak is the strongest and the $\nu_2$ ($B_g$) peak the weakest[19].

According to position shift and intensity change of Raman peaks with increasing pressure in 0-4.3GPa as show in Fig. 1, it is clear that $\nu_1$ and $\nu_3$ peaks were more sensitive to pressure than $\nu_2$ and $\nu_4$ peaks. In this pressure range, both position and intensity almost unchanged for $\nu_2$ and $\nu_4$ peaks, except for the gradually disappearance of $\nu_2$ ($B_g$) peak when pressure reached to 2.1GPa. However, the shift of $\nu_1$ and $\nu_3$ peaks at the same pressure range was very obviously, and reached an average rate of 5.29cm$^{-1}$GPa$^{-1}$ and 5.22 cm$^{-1}$GPa$^{-1}$, respectively. Moreover, the intensity of $\nu_1$ and $\nu_3$ peaks, especially for $\nu_1$ peaks, decreased quickly with pressure raising. These indicate that pressure effect is more sensitive for Cr-O stretching modes than for O-Cr-O bending modes in zircon-type CaCrO$_4$. When pressure reached 6.0GPa, a new Raman peak appeared at 878cm$^{-1}$(marked with $\nu_5$), and the intensity of other peaks distinctly declined accompanying with the broadening of $\nu_4$ (B$_g$) peak, as showed in Fig. 2. Clearly, a new crystal structure has begun to appear at 6.0GPa. When pressure continuously increased to 7.7GPa, two other new Raman peaks were observed at 337cm$^{-1}$($\nu_1$) and 920cm$^{-1}$($\nu_6$), respectively, while $\nu_4$ (A$_g$) peak vanished. In addition, the intensity of $\nu_2$ (A$_g$), $\nu_1$(A$_g$) and $\nu_3$ (B$_g$/E$_g$) peaks reduced quickly, and $\nu_4$ (B$_g$) peak sequentially broadened, tending to split into two peaks. Apparently, two different kinds of crystal structure coexisted in the pressure range from 6.0GPa to 7.7GPa in CaCrO$_4$. One crystal phase is the zircon-type structure, and the other is the new high-pressure phase, which we will discuss likely to be a scheelite-type structure in the following context. With pressure increasing to 10.5GPa, $\nu_1$ and $\nu_3$ peaks vanished entirely and other two new peaks came forth at 372cm$^{-1}$($\nu_2$) and 951cm$^{-1}$($\nu_7$). At the same time, $\nu_4$ (B$_g$) peak has split into two new peaks at 386cm$^{-1}$($\nu_3$) and 404cm$^{-1}$ ($\nu_4$). Heretofore, all the Raman peaks of zircon-type structure disappeared. This suggested the structural phase transition in CaCrO$_4$ under high pressure has been finished completely. With pressure further increasing until to the highest 26.2GPa reached in this experiment, no other new Raman peak was observed, except that Raman peaks of the high-pressure new phase become more broaden with increasing pressure. This is possibly an indication that the sample even trended to be amorphous at 26.2GPa. Finally, we released the pressure to lowest (about 0.1GPa) and found the Raman spectrum just remained the peaks of high-pressure structure. Moreover, this character unchanged 10 days later. This means the phase transition started at near 6GPa is nonreversible and the high-pressure phase can be kept steadily as a metastable phase at ambient conditions.

From the pressure dependence of Raman peaks showed in Fig. 3, four pressure regions with different structural characters were observed for $CaCrO_4$ under high pressure. In Region  there only existed zircon-type structure. Cr-O stretching modes exhibited more active to pressure than O-Cr-O bending modes in this region. This was illustrated by the shift rate of phonons frequencies and intensity change of Raman peaks. Zircon-type and new high-pressure structure coexisted in Region , which presented the continuous structural transition process. New phase began to appear and became stronger and stronger with increasing pressure, while zircon-type structure became more and more faint and was gradually replaced by the new high-pressure phase in this region. In region   , zircon-type structure completely disappeared and only the new high-pressure structure was saved. The intensity of Raman peaks in lower frequency region was feeble compared with high frequency region, but still discernable. In region   , Raman peaks sharply broaden and the sample tended to be amorphous. These interesting structural characters at various pressure regions have been confirmed by our synchrotron radiation experiments[21].

Now we discuss the phase relation with respect to pressure by comparison with $RVO_4$ (R=Y, Tb and Dy) system. At ambient conditions, $CaCrO_4$ and $RVO_4$ have equal zircon-type crystal structure (tetragonal, z=4), space group (No. 141, $I4_1/amd$), the same legend d electronic distribution ($3d^0 4s^0$ for $V^{5+}$ and $Cr^{6+}$) as well as similar lattice parameters. Therefore, it is reasonable to expect the comparable high-pressure behavior of $CaCrO_4$ and $RVO_4$. A. Jayaraman et al. have studied the Raman spectra of $RVO_4$ under high pressure and found the nonreversible pressure-induced structural phase transitions from zircon-type to scheelite-type in $RVO_4$ based on zero-pressure XRD of the pressure-released sample[11-12]. Their analysis showed the symmetry declined and the space group transformed to $I4_1/a$ when the structural transition happened. Recently, X. Wang et al. testified the result of structural phase transition in $YVO_4$ using *in-situ* high-pressure angle-resolved XRD[13]. It is the usual rule that there is an unambiguous decrease of V-O stretching modes frequencies with the transition from zircon-type to scheelite-type structure[11, 13]. Therefore, as a symbol of this phase transition, there must bring some new Raman peaks whose frequencies are lower than the corresponding frequency of $\nu_1$(A$_g$), $\nu_3$ (B$_g$), or $\nu_3$(E$_g$) peak. Considering our results, there was an observable decline of the Cr-O stretching modes frequencies when the structural phase transition happened, which was distinct as shown in Fig.3. For example, a new Raman peak ($\nu_5$) appeared at 878cm$^{-1}$ at 6.0GPa, which visibly less than the frequency (911cm$^{-1}$) of $\nu_1$(A$_g$) peak; another new Raman peak ($\nu_6$) appeared at 920cm$^{-1}$, which is also less than the frequency (936cm$^{-1}$) of the degenerate $\nu_3$ (A$_g$) (B$_g$/E$_g$) peak. The appearance of these new Raman peaks indicated the change of Cr-O bond strength in CrO$_4$ tetrahedra. According to the expression for force constant $\nu$: $\nu = 1/2p\sqrt{k/m}$ ($\nu$ and m are the vibrational frequency and renormalized mass (1/m=1/m$_{Cr}$+1/m$_O$), respectively), we estimated that the Cr-O bond strength decreased 7% with the decline of Cr-O bond stretching frequency from 911cm$^{-1}$ to 878cm$^{-1}$ when the structural phase transition started to happen at 6GPa. This suggested that the Cr-O bond length would increase with the appearance of new phase, as observed in YVO$_4$ phase transition[13]. So we regarded that the density increase in the structural transition result not from a volume reduction of CrO$_4$ tetrahedra, but from a more efficient packing of the coordination polyhedra and the elimination of a structural hole in the zircon-type $CaCrO_4$, as mentioned in Ref. 22 and 23. Accordingly, our Raman scattering results well agree with the characters of zircon-type to scheelite-type structural transition for $RVO_4$ system under high pressure. We therefore assign the new high-pressure phase in $CaCrO_4$ observed in our experiment to a scheelite-type structure with space group $I4_1/a$ (No. 88). Certainly, the accurate phase transition and the structural determination should be decided by the high-resolution angle-resolved XRD under high pressure.

In conclusion, we investigated the pressure-dependent Raman scattering on $CaCrO_4$ in the pressure range 0-26GPa at ambient temperature. A nonreversible high-pressure phase was observed at 6GPa. The pressure dependence of Raman peaks showed a series of interesting structural transition in $CaCrO_4$ under high pressure. We analyzed these pressure-related structural characters and assigned the phase transition route was from zircon- to scheelite-type structure accompanying with an increase of Cr-O bond length in $CrO_4$ tetrahedron.

Acknowledgement:

This work was partially supported by the national nature science foundation of China, the state key fundamental research project (2002CB613301), and Chinese Academy of Sciences.

**Figure captions:**

Figure 1: Raman spectra of zircon-type $CaCrO_4$ in the pressure range from 0 to 4.3GPa.

Figure 2: Some representative Raman spectra that show the structural transition of $CaCrO_4$ under pressure to 26.2GPa.

Figure 3: Pressure dependence of the Raman peaks in $CaCrO_4$. Solid dots and stars present the frequencies of Raman peaks at various pressures for zircon-type and new high-pressure phase, respectively.  ,  , and  show the four different pressure regions related to different structural characters (see text).

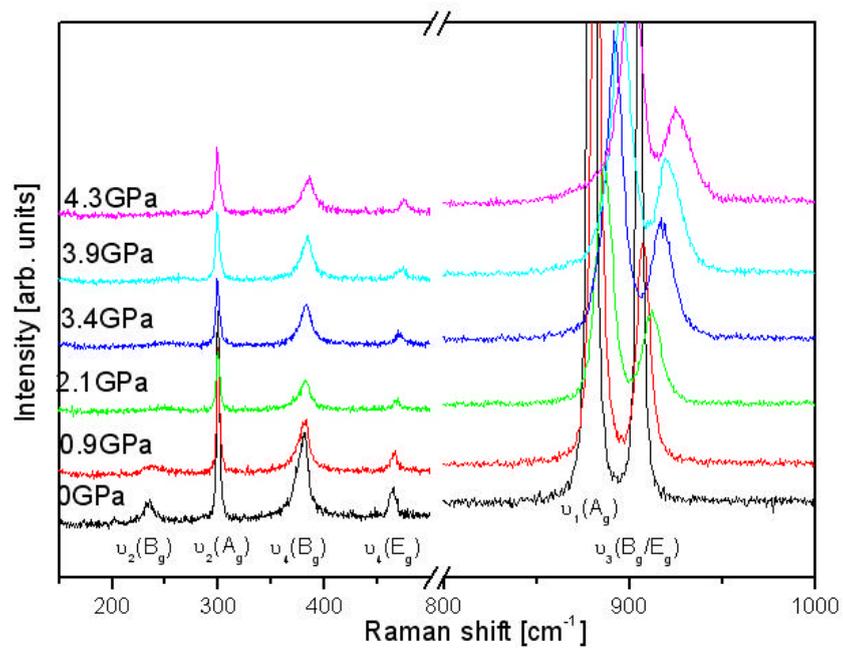

Figure 1.

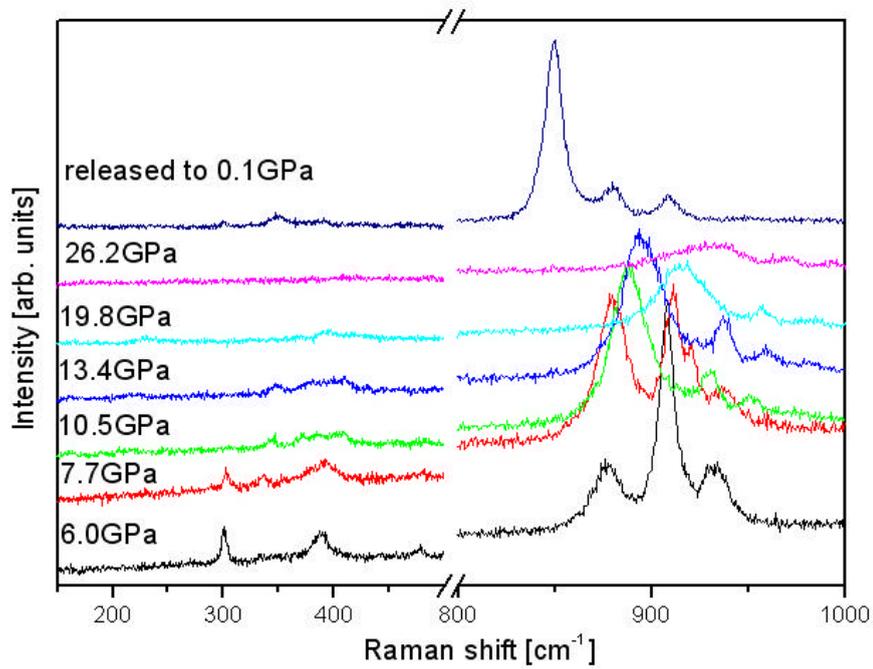

Figure 2.

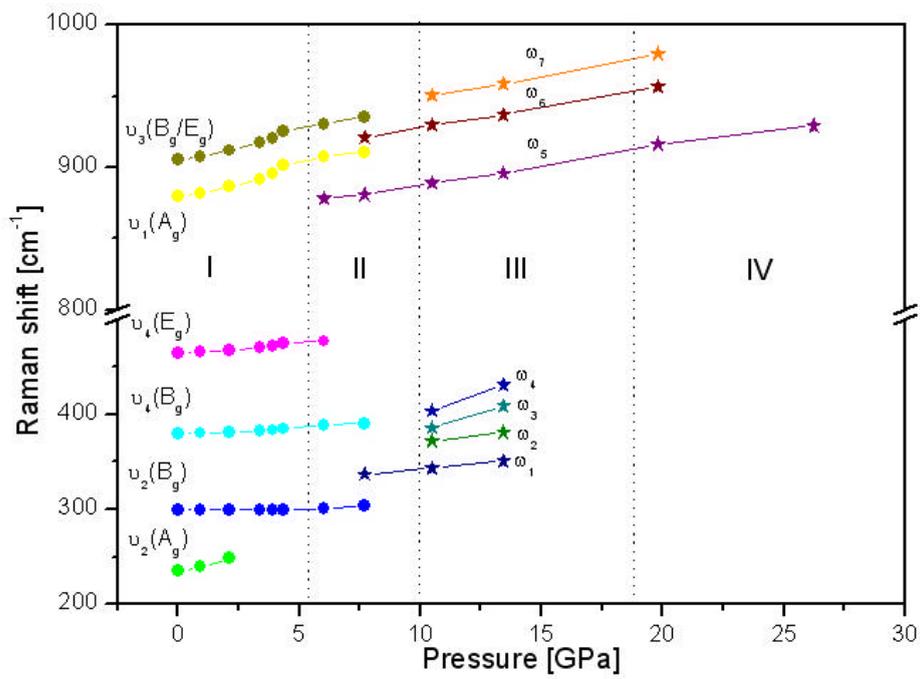

Fig. 3.